\documentclass[traditabstract]{aa} % for the abstract without structuration 
\usepackage{graphicx}
\usepackage{txfonts}
\begin{document}

\def\etal{{et al. \rm}}
\def\teff{$T_\mathrm{eff}$}
\def\logg{$\log g$}
\def\micro{$\xi$}
\def\kms{km s$^{-1}$}
\def\p{$\pm$}
\def\fe{[Fe/H]}
\def\c{{\it CoRoT}}

   \title{Abundance study of the two solar-analogue \c \ targets HD 42618 and HD 43587 from HARPS spectroscopy\thanks{Based on observations collected at the La Silla Observatory, ESO (Chile) with the HARPS spectrograph at the 3.6-m telescope, under programme LP185.D-0056.}}
   \titlerunning{An abundance study of the solar-analogue {\it CoRoT} targets HD 42618 and HD 43587}
   \authorrunning{T. Morel et al.}

      \author{T. Morel
          \inst{1}
          \and
          M. Rainer
          \inst{2}
          \and
          E. Poretti
          \inst{2}
          \and
          C. Barban
          \inst{3}
          \and
          P. Boumier
          \inst{4}
          }

   \offprints{Thierry Morel, \email{morel@astro.ulg.ac.be}.}
   \institute{
   Institut d'Astrophysique et de G\'eophysique, Universit\'e de Li\`ege, All\'ee du 6 Ao\^ut, B\^at. B5c, 4000 Li\`ege, Belgium
   \and
   INAF -- Osservatorio Astronomico di Brera, via E. Bianchi 46, 23807 Merate (LC), Italy
   \and
   LESIA, CNRS, Universit\'e Pierre et Marie Curie, Universit\'e Denis Diderot, Observatoire de Paris, 92195 Meudon Cedex, France
   \and
   Institut d'Astrophysique Spatiale, UMR8617, Universit\'e Paris XI, B\^atiment 121, 91405 Orsay Cedex, France 
   }

   \date{Received 11 December 2012 / Accepted 11 February 2013}

   \abstract{We present a detailed abundance study based on spectroscopic data obtained with HARPS of two solar-analogue main targets for the asteroseismology programme of the {\it CoRoT} satellite: HD 42618 and HD 43587. The atmospheric parameters and chemical composition are accurately determined through a fully differential analysis with respect to the Sun observed with the same instrumental set-up. Several sources of systematic errors largely cancel out with this approach, which allows us to narrow down the 1-$\sigma$ error bars to typically 20 K in effective temperature, 0.04 dex in surface gravity, and less than 0.05 dex in the elemental abundances. Although HD 42618 fulfils many requirements for being classified as a solar twin, its slight deficiency in metals and its possibly younger age indicate that, strictly speaking, it does not belong to this class of objects. On the other hand, HD 43587 is slightly more massive and evolved. In addition, marked differences are found in the amount of lithium present in the photospheres of these two stars, which might reveal different mixing properties in their interiors. These results will put tight constraints on the forthcoming theoretical modelling of their solar-like oscillations and contribute to increase our knowledge of the fundamental parameters and internal structure of stars similar to our Sun.}


   \maketitle
%
%________________________________________________________________

\section{Introduction} \label{sect_introduction}
Space-borne observatories dedicated to asteroseismology and the detection of planetary transits, such as the {\it Kepler} (Borucki \etal \cite{borucki10}) or \c \ (Baglin \etal \cite{baglin09}) satellites, are providing a wealth of very high quality photometric observations. Modelling these data places very strong constraints on stellar models and is currently leading to great advances in our understanding of a variety of physical processes that take place in stars with widely different properties and evolutionary status. In particular, observations of solar-like oscillations in late-type, main-sequence stars can be used to determine some of their fundamental stellar parameters with unprecedented accuracy or to gain important insights into their internal structure (e.g., Christensen-Dalsgaard \& Houdek \cite{christensen_dalsgaard_houdek10}, and references therein). Despite these very promising prospects and the dramatic achievements that have already been made, a good knowledge of the quantities that are not accessible from an asteroseismic analysis (most notably the effective temperature and the chemical composition) remains an essential ingredient of any successful theoretical modelling and is needed for trustworthy inferences about the star properties (e.g., Carrier \etal \cite{carrier05}; Basu \etal \cite{basu10}). In the case of \c, in particular, the photometric observations have been supported by ground-based spectroscopic ones since the launch date in order to provide these quantities (see Poretti \etal \cite{poretti13}).

This paper is an effort to obtain such accurate constraints and restrict the range of possible seismic models for two bright, solar-analogue main targets for the asteroseismology programme of the \c \ space mission: \object{HD 42618} (HIP 29432; G3 V, $V$ = 6.84 mag) and HD 43587A (HR 2251, HIP 29860, GJ 231.1A; G0 V, $V$ = 5.70 mag). The latter (hereafter HD 43587 for simplicity) is the primary of a quadruple system with a pair of distant, common-proper-motion components and a closeby M-type companion (e.g., Fuhrmann \cite{fuhrmann11}). These two nearby stars are located about 20 pc from the Sun and lie towards the anticentre direction of the Galaxy ($l$ $\sim$ 203$\degr$ and $b$ $\sim$ --5$\degr$). They have been observed by \c \ during long runs (LRa04/LRa05 for HD 42618 and LRa03 for \object{HD 43587}) with a typical duration of about 150 days. 

The extensive observations of these two stars by \c \ offer a rare opportunity to examine in some detail the internal properties of early, main-sequence G stars and to evaluate to what extent they differ from those prevailing in the Sun. This is of particular interest in the case of HD 42618, which has been recognized as a close solar analogue (e.g., Soubiran \& Triaud \cite{soubiran_triaud04}) and is, as we argue below, the star most similar to our Sun in terms of fundamental parameters ever observed by the satellite in the course of its asteroseismology programme. On the other hand, the binarity of HD 43587 provides supplementary information, such as the dynamical mass, which can be compared with the asteroseismic estimate. The full analysis of the \c \ data will be presented for HD 42618 by Barban \etal (in preparation) and for HD 43587 by Boumier \etal (in preparation). A preliminary report of the results for HD 42618 can be found in Barban \etal (\cite{barban13}).

\begin{figure*}[t]
\centering
\includegraphics[width=15cm]{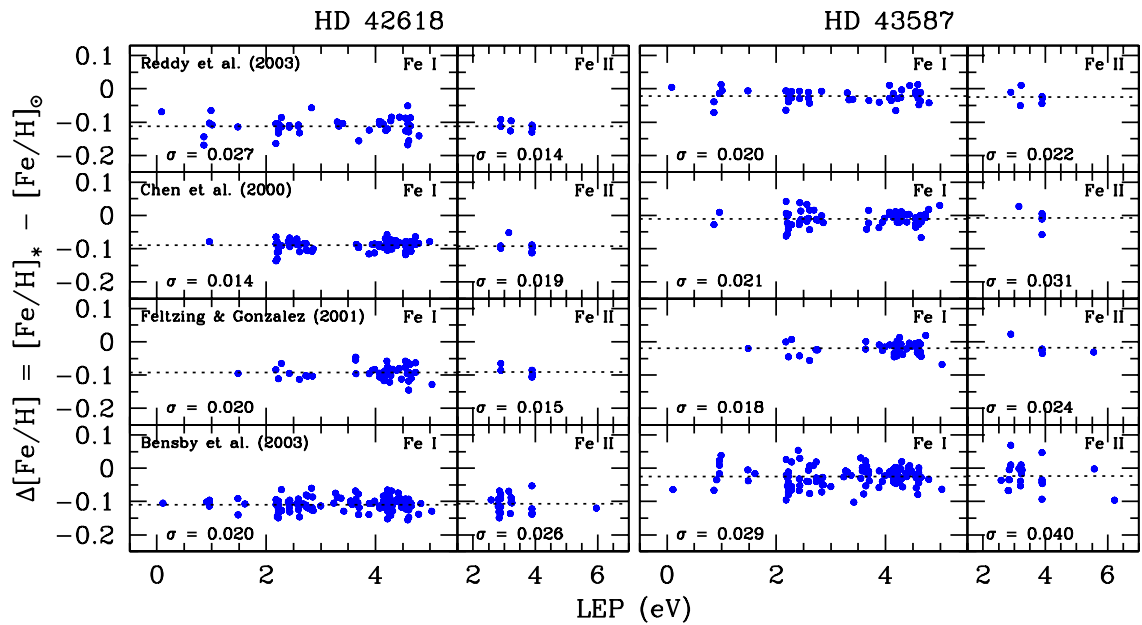}
\caption{Differences on a line-to-line basis for each line list between the abundances derived from the \ion{Fe}{i} and \ion{Fe}{ii} lines in the \c \ targets and in the Sun as a function of the lower excitation potential. The results for HD 42618 and HD 43587 are shown in the left-hand and right-hand panels, respectively. The mean iron abundance obtained for each line list is shown as a dashed line. The line-to-line scatters (in dex) are indicated.}
\label{fig_Fe_EP}
\end{figure*}

This paper is organised as follows: Section \ref{sect_observations} discusses the observations and data reduction. The methods of analysis used are described in Sect.\ref{sect_methods}. The results are presented in Sect.\ref{sect_results} and their reliability is examined in Sect.\ref{sect_reliability}. Section \ref{sect_discussion} is devoted to a discussion of the abundance pattern of our two targets with respect to that of the Sun. Finally, our main results and conclusions are summarized in Sect.\ref{sect_conclusion}.

\section{Observations and data reduction} \label{sect_observations}
Three high-resolution, optical spectra of each \c \ target were acquired during the period December 2010--January 2011 in the framework of the ESO large programme LP185.D-0056 (Poretti \etal \cite{poretti13}) with the HARPS spectrograph attached to the 3.6-m telescope at La Silla Observatory (Chile). In the high-efficiency (EGGS) mode, the spectral range covered is 3780--6910 \AA \ and the resolving power $R$ $\sim$ 80\,000. The signal-to-noise ratio (S/N) at 5815 \AA, as measured on each exposure, lies in the range 263--275 for HD 42618 and 287--392 for HD 43587. The data reduction (i.e., bias subtraction, flat-field correction, removal of scattered light, order extraction, merging of the orders, and wavelength calibration) was carried out using dedicated tools developed at Brera observatory. The spectra were subsequently put in the laboratory rest frame and were continuum-normalised by fitting low-order cubic spline or Legendre polynomials to the line-free regions using standard tasks implemented in the IRAF\footnote{{\tt IRAF} is distributed by the National Optical Astronomy Observatories, operated by the Association of Universities for Research in Astronomy, Inc., under cooperative agreement with the National Science Foundation.} software. The individual exposures were co-added (weighted by the S/N) to create an averaged spectrum, which was subsequently used for the abundance analysis. 

A series of five consecutive solar exposures (with a typical S/N of $\sim$350) were obtained on 21 December 2011 with the same instrumental configuration. Owing to the faintness of the observable asteroids at the time of observation, daytime-sky spectra were obtained. The telescope was pointed as close as possible to the Sun position to minimize the effect on the line profiles of scattering in the Earth's atmosphere (Gray \etal \cite{gray00}). The impact on our results of this possible mismatch between our skylight spectrum and the true solar one is examined in Sect.\ref{sect_results}. Exactly the same reduction steps as above were applied and the weighted average spectrum was used for the solar analysis. 

\section{Methods of analysis} \label{sect_methods}
The atmospheric parameters (\teff, \logg, and microturbulence \micro) and abundances of 22 metals were self-consistently determined from the spectra using a classical curve-of-growth analysis. The Li abundance was determined from spectral synthesis. In each case, Kurucz plane-parallel atmospheric models and the 2010 version of the line analysis software MOOG originally developed by Sneden (\cite{sneden73}) were used. Molecular equilibrium was achieved taking into account the most common 22 molecules. All models were computed with opacity distribution functions (ODFs) incorporating the solar abundances of Grevesse \& Sauval (\cite{grevesse_sauval98}) and a more comprehensive treatment of molecules compared to the ODFs computed by Kurucz (\cite{kurucz90}). More details are provided by Castelli \& Kurucz (\cite{castelli_kurucz04}).\footnote{See also {\tt http://wwwuser.oat.ts.astro.it/castelli/}.} A length of the convective cell over the pressure scale height, $\alpha$ = 1.25, and no overshooting were assumed. Appropriate ODFs and Rosseland opacity tables (rounded to the nearest 0.1 dex) were used according to the mean Fe abundance derived from the spectral analysis (because the mean abundance of the $\alpha$ elements is solar within the errors, no models with enhancements of these elements were considered). 

Thanks to the similarity in terms of fundamental parameters between our targets (especially HD 42618) and the Sun, enforcing a strictly differential analysis is expected to minimize the systematic errors that arise either from the data treatment (e.g., continuum placement) or from physical effects (e.g., granulation). This approach is particularly advantageous in that it seriously mitigates the effects of departures from local thermodynamic equilibrium (LTE), which can adversely affect the accuracy of the results. Such non-LTE effects are known to become increasingly important with decreasing metallicity. As we show below, HD 42618 is slightly more metal poor than the Sun at the $\sim$0.1 dex level. However, the differential effect of departures from LTE on the derived iron abundance and parameters appears to be negligible at these levels of metal depletion (Lind \etal \cite{lind12}). To first order, our results relative to the Sun should hence not be impacted by the assumption of LTE (the same conclusion applies to the Li abundances because the non-LTE corrections are expected to be equally very small in the Sun and in our two solar analogues; Lind \etal \cite{lind09}). 

We therefore base our results on the differences on a line-to-line basis between the abundances derived in the \c \ targets and in the Sun (see, e.g., Mel\'endez \etal \cite{melendez12} for an application of this method). In particular, the model parameters (\teff, \logg, \micro, and \fe) were iteratively modified until all the following conditions were simultaneously fulfilled: (1) the differences in the \ion{Fe}{i} abundances exhibit no trend with lower excitation potential or reduced equivalent width (the logarithm of the equivalent width divided by the wavelength of the transition); (2) the mean abundance differences found using the \ion{Fe}{i} and \ion{Fe}{ii} lines are identical; and (3) the iron abundance obtained is consistent with the model value. The abundance differences were inspected and any outlying iron spectral features removed from the analysis (this procedure was also used when determining the abundances of the other chemical elements). To determine the zero-point solar abundances, we adopted \teff \ = 5777 K, \logg \ = 4.44 dex (Cox \cite{cox00}) and adjusted the microturbulence until there was no dependence between the reduced equivalent width and the \ion{Fe}{i}-based abundances. We obtain for the Sun a mean value, $\xi \sim$ 0.86 \kms, as reported in some previous works (e.g., Bensby \etal \cite{bensby03}). To assess the uncertainties associated to the exact choice of the diagnostic lines used and the line atomic data (oscillator strengths, broadening parameters), the analysis was carried out using four different line lists taken from the literature: Bensby \etal (\cite{bensby03}), Chen \etal (\cite{chen00}), Feltzing \& Gonzalez (\cite{feltzing_gonzalez01}), and Reddy \etal (\cite{reddy03}). The study of Feltzing \& Gonzalez (\cite{feltzing_gonzalez01}) concentrated on metal-rich stars and, as a result, the lines used are generally weaker than those in the other line lists. Atomic lines significantly affected by telluric features were discarded from the analysis (the telluric atlas of Hinkle \etal \cite{hinkle00} was used). The oxygen abundance could unfortunately not be determined because \ion{[O}{i]} $\lambda$6300 is affected by a strong telluric feature in our reference solar spectrum. The \ion{Fe}{i} and \ion{Fe}{ii} abundance differences are shown as a function of the excitation potential (used to constrain \teff) in Fig.\ref{fig_Fe_EP}. The typical line-to-line scatter is about 0.02 dex.

\begin{figure}[h]
\centering
\includegraphics[width=7.8cm]{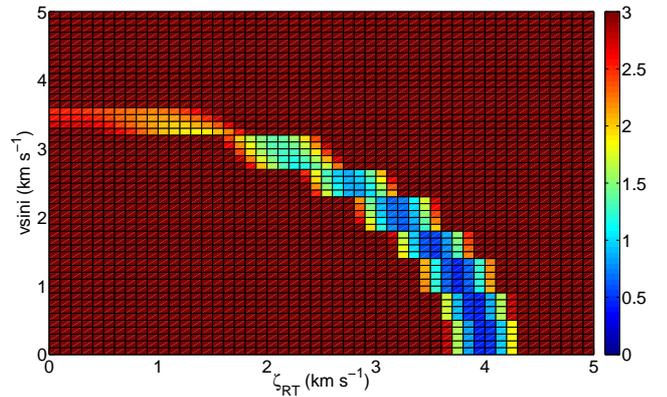}
\caption{Variation for HD 42618 of the fit quality (colour-coded as a function of $\chi^2_{\rm r}$) for different combinations of $v\sin i$ and $\zeta_\mathrm{RT}$ in the case of \ion{Fe}{i} $\lambda$6705.1. The best fit is found for $v\sin i$ = 0.9 and $\zeta_\mathrm{RT}$ = 3.7 \kms \ (and \fe \ = --0.13 dex).}
\label{fig_vsini_vmacro}
\end{figure}

\begin{figure}[h]
\centering
\includegraphics[width=7.5cm]{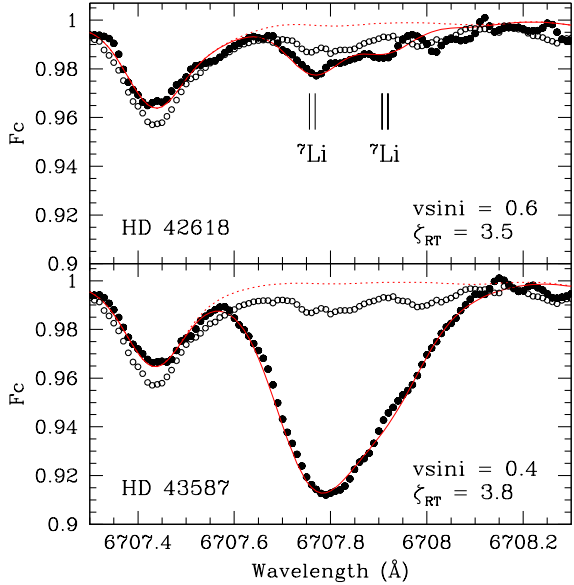}
\caption{Comparison between the observed and synthetic \ion{Li}{i} $\lambda$6708 line profiles for HD 42618 ({\it upper panel}) and HD 43587 ({\it bottom panel}). In each panel, the observed spectrum is shown ({\it filled circles}) along with the solar one for comparison ({\it open circles}). The best-fitting profiles and those assuming no lithium are shown as solid and dotted lines, respectively. The parameters have been derived using the line list of Reddy \etal (\cite{reddy03}). The $v\sin i$ and macroturbulence values adopted (in \kms) are indicated. The vertical tickmarks in the upper panel indicate the location of the $^7$Li components.}
\label{fig_li}
\end{figure}

The determination of the Li abundance relied on a spectral synthesis of \ion{Li}{i} $\lambda$6708. The lithium abundance was determined using the accurate laboratory atomic data of Smith \etal (\cite{smith98}). A solar isotopic ratio $^7$Li/$^6$Li = 12.18 was adopted (Asplund \etal \cite{asplund09}). The van der Waals damping parameters for the lithium components were taken from Barklem \etal (\cite{barklem00}). Spectral features of the CN, TiO, MgH, and SiH molecules were considered (more details can be found in an upcoming abundance study of the red giants in the \c \ seismology fields; Morel et al., in preparation). A knowledge of the total amount of line broadening is needed to perform the synthesis. An attempt was therefore made to separate the contribution of stellar rotation and radial-tangential macroturbulence, $\zeta_\mathrm{RT}$, through fitting three relatively unblended \ion{Fe}{i} lines, two of which lie in the vicinity of the Li doublet (\ion{Fe}{i} $\lambda$6265.1, 6703.6, and 6705.1 \AA). The other free parameter was \fe \ (for all stars, the iron abundances found are identical, within the errors, with the mean values derived from the curve-of-growth analysis). Instrumental broadening was determined from calibration lamps and assumed to be Gaussian with a full width at half-maximum (FWHM) of 0.079 \AA \ at the wavelength of the Li feature. Values in the range 0.4--0.9 and 3.1--4.2 \kms \ were inferred for $v\sin i$ and $\zeta_\mathrm{RT}$, respectively. Although such macroturbulent velocities are typical of solar-like dwarfs (see fig.3 of Valenti \& Fischer \cite{valenti_fischer05}), there is a strong degeneracy in the determination of these two quantities, as shown in Fig.\ref{fig_vsini_vmacro}. Furthermore, the same analysis carried out on the solar spectrum provides a formal $v\sin i$ value that is significantly lower than the true one (we obtain $v\sin i$ = 0.6 and $\zeta_\mathrm{RT}$ = 3.8 \kms). Although this known value is within the range of acceptable solutions defined by the strip in the ($v\sin i$-$\zeta_\mathrm{RT}$) plane, this shows that these two quantities are poorly constrained because of degeneracy problems (see also, e.g., Bruntt \etal \cite{bruntt10}). For both stars, only an upper limit on $v\sin i$ of about 3 \kms \ could be established. The Fourier analysis of the spectral lines combined with the availability of dedicated libraries of synthetic spectra may allow one to separate the contribution of rotation and macroturbulence (e.g., Gray \cite{gray84}). Here, exploratory calculations of the mean profile Fourier transform suggest, based on the relative height and shape of the sidelobes (in particular the squeezing of the first one; Gray \cite{gray05}), that macroturbulence is the dominant broadening mechanism. This qualitatively supports the results of the line-profile fitting described above. Valenti \& Fischer (\cite{valenti_fischer05}) obtained $v\sin i$ = 2.1 \kms \ for HD 42618 and 2.0 \kms \ for HD 43587 adopting a macroturbulence derived from their calibration as a function of the effective temperature. As they discussed, however, their values may be overestimated for \teff \ $\gtrsim$ 5800 K. The difficulties encountered when trying to constrain this parameter from the spectroscopic material routinely secured for abundance studies is illustrated by the fact that Fuhrmann (\cite{fuhrmann04}) estimated $v\sin i$ = 1.0 \kms \ only for HD 42618 by also adopting a calibrated macroturbulent velocity ($\zeta_\mathrm{RT}$ = 3.3 \kms \ in that particular case). A low-rotation rate is consistent with their low-activity levels (Hall \etal \cite{hall09}; Gray \etal \cite{gray03}; Isaacson \& Fischer \cite{isaacson_fischer10}; Lubin \etal \cite{lubin10}). HD 43587 shows no evidence for activity cycles (Baliunas \etal \cite{baliunas95}) and has even been proposed to go through a Maunder-like minimum (Schr\"oder \etal \cite{schroeder12}). Our synthesis of \ion{Li}{i} $\lambda$6708 is based on the best-fitting values, but adopting any combination of $v\sin i$ and $\zeta_\mathrm{RT}$ within the statistically acceptable solutions leads to a fit quality that is not noticeably different. The iron and lithium abundances were adjusted until a satisfactory fit of the blend primarily formed by \ion{Fe}{i} $\lambda$6707.4 and the Li doublet was achieved. A close agreement was found in all cases between the abundance yielded by this weak iron line and the mean values found with the equivalent widths (EWs). The linear limb-darkening coefficients in the $R$ band were taken from Claret (\cite{claret00}). The fits to the lithium feature for the \c \ stars are shown in Fig.\ref{fig_li}. The absolute solar reference lithium abundance we obtain is $\log \epsilon$(Li) = +0.92 dex.

\section{Atmospheric parameters and abundances} \label{sect_results}
The atmospheric parameters and chemical composition of the two \c \ targets are provided in Table \ref{tab_results} for the four different line lists. The statistical errors are related to the uncertainties inherent to fulfilling the excitation and ionisation equilibrium of the iron lines (\teff \ and \logg) or constraining the \ion{Fe}{i} abundances to be independent of the line strength (\micro). To estimate the uncertainty in \teff, for instance, we considered the range over which the slope of the relation between the \ion{Fe}{i} abundance differences and the excitation potential is consistent with zero within the errors. The effect on the abundances was then examined by altering one of the parameters by its uncertainty, while keeping the other two fixed. In practice, however, the parameters of the model are intimately coupled such that changes in one of these parameters are necessarily accompanied by variations in the other two. We therefore also took into account the fact that, for instance, such an error in \teff \ would lead to different values for \logg, \micro, and ultimately for the abundances. This covariance term was estimated by varying \teff \ by its uncertainty, keeping it fixed, and then redetermining \logg \ and \micro \ as described above. Once again, the analysis was repeated using this new set of parameters to estimate the effect on the abundances. Finally, the line-to-line scatter, $\sigma_\mathrm{line}$, was also taken into account to compute the abundance errors. 

\begin{table*}
\caption{Atmospheric parameters and elemental abundances relative to solar obtained for HD 42618 and HD 43587 using the various line lists.}
\label{tab_results}
\centering
\begin{tabular}{lccccc}
\hline\hline
\multicolumn{6}{c}{HD 42618}\\
               & Reddy \etal (\cite{reddy03}) & Chen \etal (\cite{chen00}) & Feltzing \& Gonzalez (\cite{feltzing_gonzalez01}) & Bensby \etal (\cite{bensby03}) & Mean\\\hline

\teff \ [K]             &   5745\p25        &   5765\p19        &   5765\p27        &   5775\p21         &   5765\p17  \\         
\logg \ [cgs]           &   4.43\p0.07      &   4.49\p0.06      &   4.50\p0.06      &   4.50\p0.05       &   4.48\p0.04\\    
\micro \ [\kms]         &   1.15\p0.06      &   0.86\p0.06      &   1.17\p0.06      &   0.70\p0.05       &   0.95\p0.04\\ \hline   
$[$\ion{Fe}{i}/H$]$     & --0.11\p0.04 (45) & --0.09\p0.03 (82) & --0.09\p0.04 (50) & --0.11\p0.03 (120) & --0.10\p0.02\\    
$[$\ion{Fe}{ii}/H$]$    & --0.11\p0.04 (7)  & --0.09\p0.04 (7)  & --0.09\p0.04 (6)  & --0.11\p0.04 (25)  & --0.10\p0.03\\    
$[$\ion{Li}{i}/H$]$     &  +0.21\p0.07 (1)  &  +0.23\p0.07 (1)  &  +0.23\p0.07 (1)  &  +0.24\p0.07 (1)   &  +0.23\p0.06\tablefootmark{a}\\    
$[$\ion{C}{i}/Fe$]$     &  +0.01\p0.03 (2)  &       ...         &       ...         &       ...          &  +0.01\p0.03\\    
$[$\ion{Na}{i}/Fe$]$    & --0.01\p0.02 (2)  & --0.02\p0.02 (2)  & --0.03 (1)        &  +0.00\p0.03 (4)   & --0.01\p0.02\\    
$[$\ion{Mg}{i}/Fe$]$    & --0.01 (1)        &  +0.04 (1)        &       ...         &  +0.06\p0.03 (2)   &  +0.05\p0.02\\    
$[$\ion{Al}{i}/Fe$]$    &  +0.00 (1)        & --0.01 (1)        & --0.01 (1)        &  +0.01\p0.02 (2)   &  +0.00\p0.02\\    
$[$\ion{Si}{i}/Fe$]$    &  +0.01\p0.03 (6)  &  +0.01\p0.03 (10) &  +0.00\p0.03 (8)  &  +0.02\p0.03 (16)  &  +0.01\p0.02\\    
$[$\ion{Si}{ii}/Fe$]$   & --0.02 (1)        &       ...         &       ...         &       ...          & --0.02\p0.05\tablefootmark{b}\\	    
$[$\ion{S}{i}/Fe$]$     &  +0.00 (1)        &       ...         &       ...         &       ...          &  +0.00\p0.05\tablefootmark{b}\\	    
$[$\ion{Ca}{i}/Fe$]$    &  +0.01\p0.03 (5)  &  +0.01\p0.03 (14) &  +0.00\p0.04 (7)  &  +0.03\p0.04 (19)  &  +0.01\p0.02\\    
$[$\ion{Sc}{ii}/Fe$]$   &  +0.00\p0.02 (2)  &       ...         &  +0.00\p0.02 (2)  &       ...          &  +0.00\p0.02\\    
$[$\ion{Ti}{i}/Fe$]$    &  +0.01\p0.03 (7)  &  +0.02\p0.02 (5)  & --0.01\p0.05 (7)  &  +0.02\p0.03 (28)  &  +0.02\p0.02\\    
$[$\ion{Ti}{ii}/Fe$]$   &       ...         &       ...         &       ...         &  +0.04\p0.03 (11)  &  +0.04\p0.03\\    
$[$\ion{V}{i}/Fe$]$     & --0.01\p0.03 (6)  & --0.01\p0.02 (3)  &  +0.00\p0.04 (8)  &       ...          & --0.01\p0.02\\    
$[$\ion{Cr}{i}/Fe$]$    &  +0.02\p0.03 (4)  &  +0.00\p0.02 (2)  &  +0.00\p0.02 (5)  &  +0.01\p0.04 (12)  &  +0.00\p0.02\\    
$[$\ion{Cr}{ii}/Fe$]$   &       ...         &       ...         &  +0.03 (1)        &  +0.01\p0.03 (4)   &  +0.01\p0.02\\    
$[$\ion{Co}{i}/Fe$]$    &  +0.01\p0.04 (3)  &       ...         & --0.01\p0.07 (4)  &       ...          &  +0.00\p0.03\\    
$[$\ion{Ni}{i}/Fe$]$    & --0.02\p0.03 (15) & --0.02\p0.02 (16) & --0.02\p0.02 (21) & --0.01\p0.03 (45)  & --0.02\p0.02\\    
$[$\ion{Zn}{i}/Fe$]$    &  +0.03\p0.03 (2)  &       ...         &       ...         &  +0.02\p0.03 (2)   &  +0.03\p0.02\\    
$[$\ion{Sr}{i}/Fe$]$    &  +0.00 (1)        &       ...         &       ...         &       ...          &  +0.00\p0.05\tablefootmark{b}\\    
$[$\ion{Y}{ii}/Fe$]$    &  +0.00\p0.03 (4)  &       ...         &       ...         &       ...          &  +0.00\p0.03\\    
$[$\ion{Zr}{ii}/Fe$]$   & --0.04 (1)        &       ...         &       ...         &       ...          & --0.04\p0.05\tablefootmark{b}\\    
$[$\ion{Ba}{ii}/Fe$]$   & --0.05\p0.06 (2)  &  +0.00\p0.08 (3)  &       ...         &       ...          & --0.03\p0.05\\    
$[$\ion{Ce}{ii}/Fe$]$   &  +0.03\p0.02 (3)  &       ...         &       ...         &       ...          &  +0.03\p0.02\\    
$[$\ion{Nd}{ii}/Fe$]$   &  +0.03 (1)        &       ...         &       ...         &       ...          &  +0.03\p0.05\tablefootmark{b}\\    
$[$\ion{Eu}{ii}/Fe$]$   &  +0.07 (1)        &       ...         &       ...         &       ...          &  +0.07\p0.05\tablefootmark{b}\\
\hline
&&&&&\\
\hline\hline
\multicolumn{6}{c}{HD 43587}\\
                        & Reddy \etal (\cite{reddy03}) & Chen \etal (\cite{chen00}) & Feltzing \& Gonzalez (\cite{feltzing_gonzalez01}) & Bensby \etal (\cite{bensby03}) & Mean\\\hline

\teff \ [K]             &   5920\p22        &   5965\p21        &   5940\p27        &   5965\p26         &   5947\p17  \\         
\logg \ [cgs]           &   4.29\p0.07      &   4.42\p0.06      &   4.39\p0.06      &   4.38\p0.06       &   4.37\p0.04\\    
\micro \ [\kms]         &   1.29\p0.05      &   1.06\p0.06      &   1.42\p0.06      &   1.00\p0.04       &   1.16\p0.04\\\hline   
$[$\ion{Fe}{i}/H$]$     & --0.02\p0.03 (44) & --0.01\p0.04 (77) & --0.02\p0.03 (45) & --0.02\p0.04 (118) & --0.02\p0.02\\    
$[$\ion{Fe}{ii}/H$]$    & --0.02\p0.04 (6)  & --0.01\p0.05 (5)  & --0.02\p0.04 (5)  & --0.02\p0.05 (23)  & --0.02\p0.03\\    
$[$\ion{Li}{i}/H$]$     &  +1.11\p0.06 (1)  &  +1.15\p0.06 (1)  &  +1.13\p0.06 (1)  &  +1.14\p0.06 (1)   &  +1.13\p0.05\tablefootmark{a}\\    
$[$\ion{C}{i}/Fe$]$     &  +0.00\p0.03 (3)  &       ...         &  +0.04 (1)        &       ...          &  +0.01\p0.03\\    
$[$\ion{Na}{i}/Fe$]$    &  +0.00\p0.02 (2)  & --0.01\p0.02 (2)  & --0.01 (1)        &  +0.02\p0.04 (4)   &  +0.00\p0.02\\    
$[$\ion{Mg}{i}/Fe$]$    &  +0.03\p0.06 (2)  &  +0.05 (1)        &       ...         &  +0.06\p0.02 (2)   &  +0.06\p0.02\\    
$[$\ion{Al}{i}/Fe$]$    & --0.01 (1)        & --0.01 (1)        & --0.01 (1)        &  +0.00\p0.02 (2)   &  +0.00\p0.02\\    
$[$\ion{Si}{i}/Fe$]$    &  +0.00\p0.02 (6)  &  +0.01\p0.03 (10) &  +0.00\p0.03 (8)  &  +0.02\p0.03 (15)  &  +0.01\p0.02\\    
$[$\ion{Si}{ii}/Fe$]$   &  +0.04 (1)        &       ...         &       ...         &       ...          &  +0.04\p0.05\tablefootmark{b}\\	    
$[$\ion{Ca}{i}/Fe$]$    & --0.01\p0.02 (5)  &  +0.02\p0.05 (15) &  +0.00\p0.04 (9)  &  +0.04\p0.04 (19)  &  +0.00\p0.02\\    
$[$\ion{Sc}{ii}/Fe$]$   & --0.02\p0.03 (2)  &       ...         &  +0.01\p0.03 (2)  &       ...          &  +0.00\p0.02\\    
$[$\ion{Ti}{i}/Fe$]$    & --0.01\p0.03 (7)  &  +0.00\p0.04 (5)  & --0.03\p0.05 (8)  &  +0.00\p0.04 (27)  & --0.01\p0.02\\    
$[$\ion{Ti}{ii}/Fe$]$   &       ...         &       ...         &       ...         &  +0.02\p0.04 (12)  &  +0.02\p0.04\\    
$[$\ion{V}{i}/Fe$]$     & --0.01\p0.04 (6)  &  +0.00\p0.06 (3)  &  +0.00\p0.04 (8)  &       ...          &  +0.00\p0.03\\    
$[$\ion{Cr}{i}/Fe$]$    &  +0.00\p0.03 (4)  &  +0.00\p0.04 (2)  &  +0.00\p0.03 (5)  &  +0.02\p0.03 (12)  &  +0.01\p0.02\\    
$[$\ion{Cr}{ii}/Fe$]$   &       ...         &       ...         &  +0.01 (1)        &  +0.00\p0.02 (4)   &  +0.00\p0.02\\    
$[$\ion{Co}{i}/Fe$]$    &  +0.00\p0.04 (3)  &       ...         & --0.04\p0.05 (6)  &       ...          & --0.01\p0.03\\    
$[$\ion{Ni}{i}/Fe$]$    &  +0.00\p0.03 (15) &  +0.00\p0.03 (14) & --0.01\p0.03 (22) &  +0.00\p0.03 (41)  &  +0.00\p0.02\\    
$[$\ion{Zn}{i}/Fe$]$    &  +0.00\p0.02 (2)  &       ...         &       ...         & --0.01\p0.03 (2)   &  +0.00\p0.02\\    
$[$\ion{Sr}{i}/Fe$]$    & --0.02 (1)        &       ...         &       ...         &       ...          & --0.02\p0.05\tablefootmark{b}\\    
$[$\ion{Y}{ii}/Fe$]$    & --0.04\p0.03 (4)  &       ...         &       ...         &       ...          & --0.04\p0.03\\    
$[$\ion{Zr}{ii}/Fe$]$   & --0.05 (1)        &       ...         &       ...         &       ...          & --0.05\p0.05\tablefootmark{b}\\    
$[$\ion{Ba}{ii}/Fe$]$   & --0.09\p0.04 (3)  & --0.07\p0.03 (3)  &       ...         &       ...          & --0.08\p0.03\\    
$[$\ion{Ce}{ii}/Fe$]$   & --0.04\p0.03 (2)  &       ...         &       ...         &       ...          & --0.04\p0.03\\    
$[$\ion{Nd}{ii}/Fe$]$   & --0.02 (1)        &       ...         &       ...         &       ...          & --0.02\p0.05\tablefootmark{b}\\    
$[$\ion{Eu}{ii}/Fe$]$   & --0.11 (1)        &       ...         &       ...         &       ...          & --0.11\p0.05\tablefootmark{b}\\
\hline
\end{tabular}
\tablefoot{Given are the solar-normalised abundance ratios with respect to Fe (relative to hydrogen for iron and lithium). We use the usual notation: [A/B] = $[\log \epsilon($A$)-\log \epsilon($B$)]-[\log \epsilon($A$)-\log \epsilon($B$)]_{\sun}$ where $\log \epsilon$(A) = 12 + $\log [{\cal N}($A$)/{\cal N}($H$)]$ (${\cal N}$ is the number density of the species). Our final values for the \c \ targets are provided in the last column. The number in brackets gives the number of lines used for a given ion.
\\
\tablefootmark{a}{The absolute lithium abundances we obtain are $\log \epsilon$(Li) = +1.15 dex for HD 42618 and +2.05 dex for HD 43587.\\}
\tablefootmark{b}{Fixed error value.}}
\end{table*}

On the other hand, our results are also affected by three main sources of systematic errors. As already mentioned in Sect.\ref{sect_observations}, our daytime-sky spectrum is not expected to be a perfect match to the solar spectrum. First, a sky spectrum is obviously not an observation of a point-like source. A potentially more serious concern is also that the line profiles may be slightly filled in because of scattering in the atmosphere, with changes in the EWs amounting to a few per cent (e.g., Gray \etal \cite{gray00}; Molaro \etal \cite{molaro08}). To evaluate the impact on our results, we repeated the analysis after artificially increasing our EWs by 3\%, which can be regarded as a generous estimate (e.g., Ram{\'{\i}}rez \etal \cite{ramirez11}). In addition, we carried out the analysis using MARCS models (Gustafsson \etal \cite{gustafsson08}). Note that in contrast to ATLAS9 model atmospheres, MARCS models use solar abundances based on 3D hydrodynamical simulations (Grevesse \etal \cite{grevesse07}). Very minor differences are found with respect to the results from Kurucz models, which is indeed expected given the close similarity of the atmospheric structure for both sets of models in this parameter range (Gustafsson \etal \cite{gustafsson08}). Finally, estimating the best fit to the lithium line is subjective and is especially uncertain in the Sun and in HD 42618, which have weak features. We assigned a systematic uncertainty associated to this procedure, $\sigma_\mathrm{fit}$, of 0.05 and 0.04 dex for HD 42618 and HD 43587, respectively.

These various sources of errors (statistical and those arising from the mismatch of the reference solar spectrum, the choice of the atmosphere models and, in the case of Li, that associated to the fitting procedure) can be considered as independent and were quadratically summed to obtain the final uncertainties. Table \ref{tab_errors} shows an example of the error calculations. In the case of abundances based on a single line, $\sigma_\mathrm{line}$ cannot be determined and we conservatively set the total uncertainty to 0.05 dex.

The results obtained for a given quantity based on the $i$th line list, $x_i$, were weighted by their total random uncertainties, $\sigma_i$, to obtain the final values, $\overline{x}$, provided in the last column of Table \ref{tab_results} with the usual formula:

\begin{equation}
\overline{x} = {\Sigma (x_i/\sigma_i^2) \over \Sigma (1/\sigma_i^2)}.
\end{equation}

The corresponding uncertainty is computed as the quadratic sum of the statistical and systematic errors:
\begin{equation}
\sigma(\overline{x}) = \left[ {1 \over \Sigma (1/\sigma_i^2)} + \sigma^2_\mathrm{model} + \sigma^2_\mathrm{sky} + \sigma^2_\mathrm{fit} \right]^{1/2}{\rm .}
\end{equation}

\begin{table*}
\caption{Computation of the error budget for HD 42618 and the line list of Reddy \etal (\cite{reddy03}).}
\label{tab_errors}
\centering
\begin{tabular}{l|ccccccc|ccc|l}
\hline\hline
\multicolumn{1}{c}{} & \multicolumn{7}{c}{Statistical errors} & \multicolumn{3}{c}{Systematic errors} & \\
                     & $\sigma_\mathrm{line}$ & $\sigma_\mathrm{Teff}$ & $\sigma_\mathrm{logg}$ & $\sigma_\mathrm{\xi}$ & $\sigma_\mathrm{logg+\xi/Teff}$ & $\sigma_\mathrm{Teff+\xi/logg}$ & $\sigma_\mathrm{Teff+logg/\xi}$ & $\sigma_\mathrm{model}$ & $\sigma_\mathrm{sky}$ & $\sigma_\mathrm{fit}$ & \multicolumn{1}{c}{$\sigma$}\\\hline
\teff \ [K]           & ...   & 17    & ...   & ...   & ...   & 5     & 10    & 10    & 10    & ...  & \multicolumn{1}{c}{25} \\
\logg \ [cgs]         & ...   & ...   & 0.03  & ...   & 0.04  & ...   & 0.03  & 0.02  & 0.01  & ...  & 0.063\\
\micro \ [\kms]       & ...   & ...   & ...   & 0.030 & 0.025 & 0.009 & ...   & 0.030 & 0.006 & ...  & 0.051\\
\hline
$[$\ion{Fe}{i/H}$]$   & 0.027 & 0.015 & 0.000 & 0.004 & 0.011 & 0.002 & 0.003 & 0.002 & 0.015 & ...  & 0.037\\
$[$\ion{Fe}{ii}/H$]$  & 0.014 & 0.007 & 0.012 & 0.006 & 0.010 & 0.014 & 0.002 & 0.003 & 0.015 & ...  & 0.031\\
$[$\ion{Li}{i}/H$]$   & ...   & 0.02  & 0.00  & 0.00  & 0.02  & 0.00  & 0.01  & 0.02  & 0.02  & 0.05 & 0.065  \\
$[$\ion{C}{i}/Fe$]$   & 0.023 & 0.015 & 0.005 & 0.005 & 0.006 & 0.003 & 0.002 & 0.004 & 0.003 & ...  & 0.030\\
$[$\ion{Na}{i}/Fe$]$  & 0.007 & 0.005 & 0.008 & 0.003 & 0.006 & 0.009 & 0.000 & 0.001 & 0.005 & ...  & 0.018\\
$[$\ion{Mg}{i}/Fe$]$  & ...   & 0.003 & 0.012 & 0.000 & 0.008 & 0.013 & 0.005 & 0.001 & 0.007 & ...  & 0.050\tablefootmark{a} \\
$[$\ion{Al}{i}/Fe$]$  & ...   & 0.005 & 0.006 & 0.005 & 0.005 & 0.008 & 0.002 & 0.000 & 0.002 & ...  & 0.050\tablefootmark{a} \\
$[$\ion{Si}{i}/Fe$]$  & 0.019 & 0.002 & 0.005 & 0.003 & 0.008 & 0.006 & 0.001 & 0.001 & 0.002 & ...  & 0.023\\
$[$\ion{Si}{ii}/Fe$]$ & ...   & 0.023 & 0.006 & 0.002 & 0.009 & 0.005 & 0.001 & 0.003 & 0.004 & ...  & 0.050\tablefootmark{a} \\
$[$\ion{S}{i}/Fe$]$   & ...   & 0.015 & 0.003 & 0.005 & 0.007 & 0.001 & 0.001 & 0.003 & 0.000 & ...  & 0.050\tablefootmark{a} \\
$[$\ion{Ca}{i}/Fe$]$  & 0.019 & 0.010 & 0.009 & 0.001 & 0.005 & 0.010 & 0.002 & 0.002 & 0.003 & ...  & 0.027\\
$[$\ion{Sc}{ii}/Fe$]$ & 0.011 & 0.004 & 0.008 & 0.004 & 0.006 & 0.006 & 0.009 & 0.001 & 0.003 & ...  & 0.020\\
$[$\ion{Ti}{i}/Fe$]$  & 0.010 & 0.015 & 0.007 & 0.001 & 0.002 & 0.008 & 0.003 & 0.003 & 0.003 & ...  & 0.022\\
$[$\ion{V}{i}/Fe$]$   & 0.016 & 0.017 & 0.006 & 0.004 & 0.004 & 0.008 & 0.006 & 0.000 & 0.002 & ...  & 0.027\\
$[$\ion{Cr}{i}/Fe$]$  & 0.014 & 0.012 & 0.007 & 0.001 & 0.001 & 0.008 & 0.001 & 0.000 & 0.001 & ...  & 0.022\\
$[$\ion{Co}{i}/Fe$]$  & 0.028 & 0.007 & 0.005 & 0.003 & 0.002 & 0.007 & 0.002 & 0.001 & 0.001 & ...  & 0.031\\
$[$\ion{Ni}{i}/Fe$]$  & 0.023 & 0.007 & 0.005 & 0.001 & 0.003 & 0.006 & 0.000 & 0.001 & 0.001 & ...  & 0.026\\
$[$\ion{Zn}{i}/Fe$]$  & 0.018 & 0.005 & 0.002 & 0.001 & 0.007 & 0.002 & 0.004 & 0.001 & 0.001 & ...  & 0.021\\
$[$\ion{Sr}{i}/Fe$]$  & ...   & 0.015 & 0.008 & 0.003 & 0.001 & 0.008 & 0.001 & 0.001 & 0.004 & ...  & 0.050\tablefootmark{a} \\
$[$\ion{Y}{ii}/Fe$]$  & 0.023 & 0.003 & 0.006 & 0.002 & 0.005 & 0.006 & 0.004 & 0.002 & 0.004 & ...  & 0.026\\
$[$\ion{Zr}{ii}/Fe$]$ & ...   & 0.004 & 0.006 & 0.004 & 0.005 & 0.004 & 0.010 & 0.002 & 0.001 & ...  & 0.050\tablefootmark{a} \\
$[$\ion{Ba}{ii}/Fe$]$ & 0.055 & 0.001 & 0.000 & 0.010 & 0.001 & 0.002 & 0.008 & 0.003 & 0.011 & ...  & 0.058\\
$[$\ion{Ce}{ii}/Fe$]$ & 0.011 & 0.001 & 0.006 & 0.002 & 0.008 & 0.005 & 0.009 & 0.004 & 0.002 & ...  & 0.019\\
$[$\ion{Nd}{ii}/Fe$]$ & ...   & 0.001 & 0.007 & 0.004 & 0.010 & 0.005 & 0.013 & 0.003 & 0.003 & ...  & 0.050\tablefootmark{a} \\
$[$\ion{Eu}{ii}/Fe$]$ & ...   & 0.004 & 0.006 & 0.005 & 0.006 & 0.005 & 0.010 & 0.000 & 0.001 & ...  & 0.050\tablefootmark{a} \\
\hline
\end{tabular}
\tablefoot{The line-to-line scatter is given by $\sigma_\mathrm{line}$. Other statistical errors arise from uncertainties in the effective temperature ($\sigma_\mathrm{Teff}$), surface gravity ($\sigma_\mathrm{logg}$), or microturbulence ($\sigma_\mathrm{\xi}$). The changes in, for instance, \logg, $\xi$, and abundances associated to varying \teff \ by its uncertainty are noted as $\sigma_\mathrm{logg+\xi/Teff}$. As discussed in Sect.\ref{sect_results}, the systematic uncertainties are related to the choice of either Kurucz or MARCS model atmospheres ($\sigma_\mathrm{model}$), the use of a skylight solar spectrum ($\sigma_\mathrm{sky}$), or the subjectivity in estimating the best fit to the lithium feature ($\sigma_\mathrm{fit}$). The last column gives the total uncertainty.
\\
\tablefootmark{a}{Fixed error value.}}
\end{table*}

\begin{figure}[h]
\centering
\includegraphics[width=9cm]{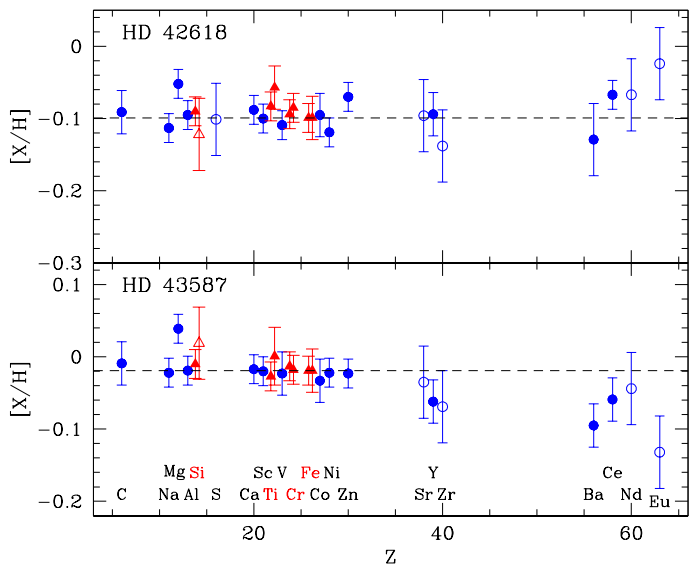}
\caption{Abundance pattern of HD 42618 ({\it upper panel}) and HD 43587 ({\it bottom panel}) relative to the Sun. The abundances of elements based on two different ionisation stages (\ion{Si}{i}/\ion{Si}{ii}, \ion{Ti}{i}/\ion{Ti}{ii}, \ion{Cr}{i}/\ion{Cr}{ii}, and \ion{Fe}{i}/\ion{Fe}{ii}) are shown as triangles. For clarity, the points have been slightly shifted along the abscissa axis in this case. The open symbols denote abundances based on a single spectral line, whose uncertainty has been conservatively set to 0.05 dex. The dashed line shows the mean [Fe/H] abundance.}
\label{fig_patterns}
\end{figure}

For HD 42618, we obtain: \teff \ = 5765\p17 K, \logg \ = 4.48\p0.04 dex and \fe \ = --0.10\p0.02 dex, while HD 43587 is slightly hotter (in accordance with its spectral type) and metal rich:  \teff \ = 5947\p17 K, \logg \ = 4.37\p0.04 dex and \fe \ = --0.02\p0.02 dex. As shown in Fig.\ref{fig_patterns}, our gravity estimates are supported for the two targets by the fact that ionisation balance is fulfilled within the errors for three other elements (Si, Ti, and Cr). The mean abundance of these elements and iron discussed in the following is the weighted (by the inverse variance) average of the values for the two ions. The chemical mixture is in both cases close to solar, except perhaps for a slight magnesium excess in the two targets and a slight deficiency of some rare, heavy species (e.g., the neutron-capture elements Ba and Eu) in HD 43587 (Fig.\ref{fig_patterns}). However, the abundance of these elements is based on a few lines that are generally weak and caution should be exercised. The star HD 42618 is depleted in lithium to similar levels as in the Sun, whereas HD 43587 is dramatically more lithium-rich by about one order of magnitude. However, its surface Li abundance is by no means particularly high when compared to stars with similar temperatures, metallicities, and ages (e.g., Sestito \& Randich \cite{sestito_randich05}; Ram{\'{\i}}rez \etal \cite{ramirez12}). Its beryllium abundance is roughly what is expected on the basis of its lithium content (Boesgaard \etal \cite{boesgaard01}). The star HD 43587 is actually part of a quadruple system with not only a pair of distant low-mass components (HD 43587B and C; Pravdo \etal \cite{pravdo06}), but also a relatively close ($\rho$ $\lesssim$ 1\arcsec) and much dimmer M-type companion, HD 43587Ab, following a long ($\cal{P} \sim$ 30 yr) and very eccentric ($e \sim$ 0.8) orbit (Catala \etal \cite{catala06}; Vogt \etal \cite{vogt02}; Hartkopf \etal \cite{hartkopf12}; Katoh \etal \cite{katoh13}). As discussed by Fuhrmann (\cite{fuhrmann11}), neglecting the contamination of the integrated spectrum by this cooler component may lead to a small error in \teff, \logg \ and \fe \ of about 20 K, 0.04 dex and 0.02 dex. We therefore cannot rule out that our values are underestimated by similar amounts.\footnote{Our spectrum is contaminated by the presence of the companion because the fibre entrance aperture projected on the sky of HARPS in the EGGS mode is 1.4\arcsec, while interferometric observations -- which were obtained at the end December 2010 and are therefore nearly contemporaneous with our spectroscopy -- give a separation between the two components of about 0.88\arcsec \ (Hartkopf \etal \cite{hartkopf12}).}

\section{Reliability of the results} \label{sect_reliability}
\subsection{Fundamental parameters} \label{sect_reliability_parameters}
\begin{figure}[h]
\centering
\includegraphics[width=8.5cm]{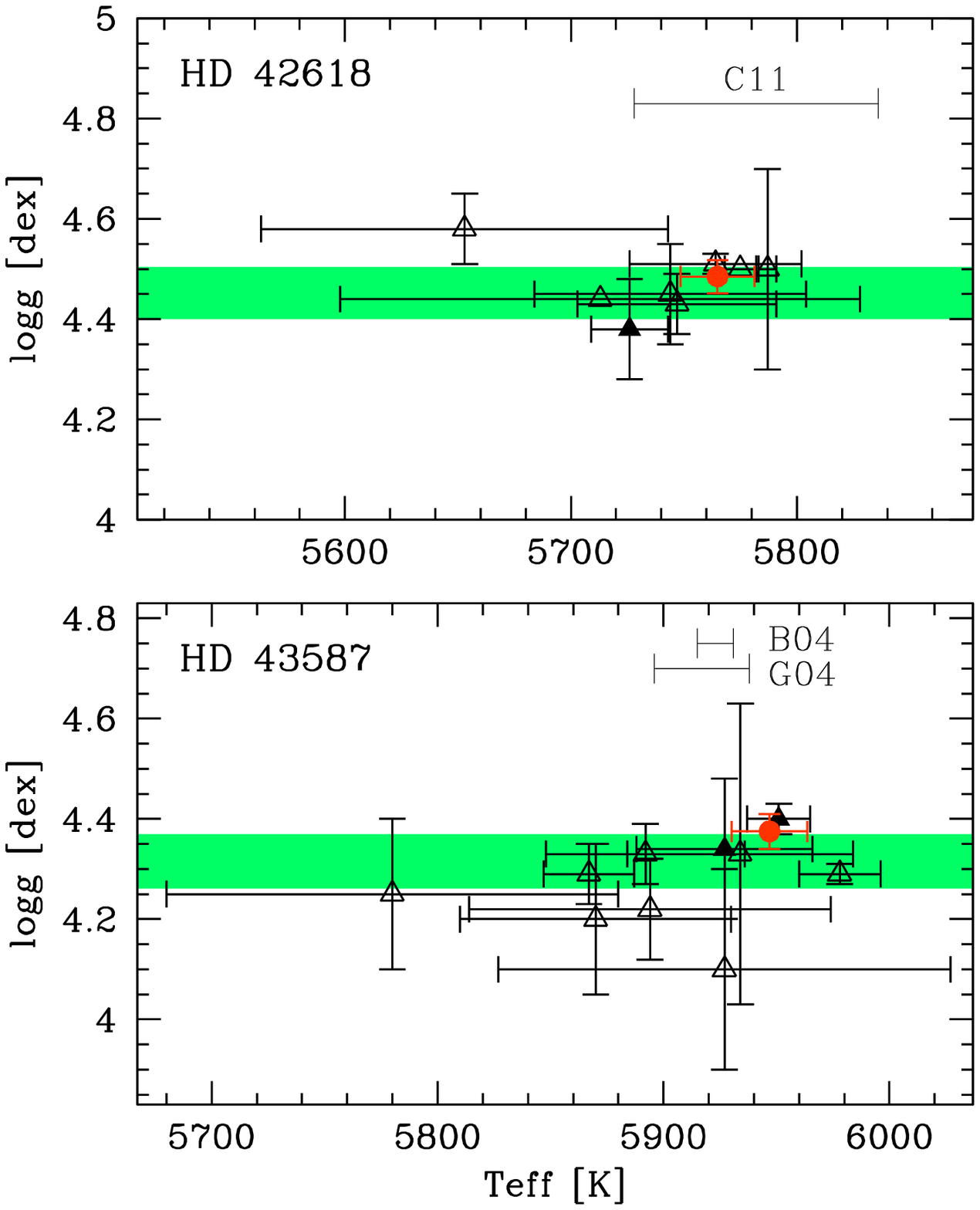}
\caption{Comparison between our \teff \ and \logg \ values ({\it filled circles}) and those in the literature (Boesgaard \etal \cite{boesgaard01}; Bruntt \etal \cite{bruntt04}; da Silva \etal \cite{da_silva11}; Fuhrmann \cite{fuhrmann04}, \cite{fuhrmann11}; Gillon \& Magain \cite{gillon_magain06}; Gray \etal \cite{gray03}; Kovtyukh \etal \cite{kovtyukh04}; Mishenina \etal \cite{mishenina04}, \cite{mishenina08}; Neves \etal \cite{neves12}; Ram{\'{\i}}rez \etal \cite{ramirez13}; Reddy \etal \cite{reddy03}; Takeda \etal \cite{takeda07}; Valenti \& Fischer \cite{valenti_fischer05}). The values obtained for HD 43587 by Bruntt \etal (\cite{bruntt04}) and Gillon \& Magain (\cite{gillon_magain06}) using different methods are plotted separately. {\it Filled triangles}: parameters derived from excitation and ionisation balance of the iron lines, {\it open triangles}: parameters derived using different techniques. Also shown are the results of Casagrande \etal (\cite{casagrande11}) for HD 42618 based on colour indices (C11) and those for HD 43587 of Bruntt \etal (\cite{bruntt04}; B04) and Gray (\cite{gray94}; G04) based on line-depth ratios. The horizontal strip corresponds to the 1-$\sigma$ error bar for the surface gravity determined from seismic scaling relations (see Sect.\ref{sect_reliability_parameters}).}
\label{fig_teff_logg}
\end{figure}

As seen in Fig.\ref{fig_teff_logg}, our parameters agree well with the most precise results available in the literature. Our \teff \ value for HD 42618 is identical, within the error bars, with that of Casagrande \etal (\cite{casagrande11}) based on colour indices (5782\p54 K). Our value for HD 43587 also agrees well with the results of Gray (\cite{gray94}) and Bruntt \etal (\cite{bruntt04}) who used line-depth ratios (5917\p21 and 5923\p8 K, respectively).
 
A seismic scaling relation using the frequency of maximum oscillation power, $\nu_{\rm max}$, can also be used to precisely infer the surface gravity (e.g., Brown \etal \cite{brown91}):

\begin{equation}
\log g = \log g_\odot + \log \left(
{{\nu_{\rm max} \over \nu_{{\rm max}, \, \odot}}}
\right)
+ {1 \over 2}
\log \left({T_{\rm eff}\over {\rm T}_{{\rm eff}, \, \odot}}\right) {\rm .}
\label{eq:loggnumax}
\end{equation}

This relation is largely insensitive to the effective temperature assumed ($\Delta T_{\rm eff}$ = 100 K leads to $\Delta \log g$ $\sim$ 0.004 dex only for Sun-like stars). The seismic and spectroscopic gravities can therefore be regarded as being virtually completely independent. Our effective temperatures, $\nu_{{\rm max}, \, \odot}$ = 3100 $\mu$Hz, and the preliminary $\nu_{\rm max}$ values obtained from the analysis of the \c \ data for HD 42618 (3200\p100 $\mu$Hz; Barban \etal \cite{barban13}) and HD 43587 (2300\p100 $\mu$Hz; Boumier et al., in preparation) were assumed to compute the seismic gravities. Propagating the errors in these quantities leads to very small uncertainties in \logg \ (of the order of 0.01--0.02 dex). However, it is very likely that these values are unrealistically low and that a systematic error related to the scaling relation itself dominates. We conservatively adopted a systematic uncertainty of 0.05 dex in the following, although the reliability of the relations that link the stellar parameters and the seismic observables has been so far validated by a number of empirical tests (e.g., Miglio \cite{miglio12}). In particular, there is no evidence for systematic differences from a statistical point of view with the gravities obtained using classical methods (e.g., the absolute difference for nearby dwarfs with respect to gravities determined from isochrone fitting is $\Delta \log g$ = 0.006\p0.065 dex; Morel \& Miglio \cite{morel_miglio12}). We finally obtained \logg \ = 4.45\p0.06 and 4.32\p0.06 dex for HD 42618 and HD 43587, respectively. As seen in Fig.\ref{fig_teff_logg}, the spectroscopic gravities and those based on scaling relations using $\nu_{{\rm max}}$ agree within the errors. A more detailed modelling of the pulsation frequencies (additionally using our \teff \ as input; see Metcalfe \etal \cite{metcalfe12}) leads for HD 42618 to a nearly identical seismic \logg \ value (Barban \etal \cite{barban13}). The agreement is, however, less gratifying for HD 43587, with a spectroscopic estimate higher at the 0.05 dex level (Boumier et al., in preparation). A similar discrepancy is observed when comparing with evolutionary tracks (see below) and might be related to the presence of the companion.

A Bayesian estimation through isochrone fitting based on the evolutionary tracks of Girardi \etal (\cite{giradi00}) was also performed with PARAM (see da Silva \etal \cite{da_silva06})\footnote{See also {\tt http://stev.oapd.inaf.it/cgi-bin/param.}} to estimate the fundamental parameters. The input values were \teff, \fe, the parallaxes and the magnitudes in the $V$ band (reddening was neglected). The results are shown in Table \ref{tab_param}. The isochrone and spectroscopic surface gravities agree within the errors for HD 42618, while the latter is significantly higher for HD 43587. As for the spectroscopic results, the presence of the faint M-type companion is also expected to slightly affect these estimates: correcting for the light contribution of the secondary (assuming $\Delta V$ $\sim$ 4.7 mag; Fuhrmann \cite{fuhrmann11}) and revising \teff \ slightly upwards by 20 K and \fe \ by 0.02 dex (Sect.\ref{sect_results}), one obtains that the star is fainter by about 0.014 mag in the $V$ band, slightly younger and more massive (Table \ref{tab_param}). The small discrepancy in \logg \ remains when taking into account that our spectroscopic value should also be revised upwards (by about 0.04 dex; Sect.\ref{sect_results}). The radius, $R$ = 1.280\p0.032 R$_{\odot}$, derived by Th\'evenin \etal (\cite{thevenin06}) for HD 43587Aa using a calibrated Barnes-Evans relation and the previous, slightly lower {\it Hipparcos} parallax is likely overestimated. On the other hand, its dynamical mass does not provide tight constraints ($M$ = 0.96\p0.18 M$_{\odot}$; Catala \etal \cite{catala06}).

\begin{table}[h]
\caption{Fundamental parameters derived using PARAM.}
\label{tab_param}
\centering
\begin{tabular}{lccc}
\hline\hline
                  & HD 42618              & HD 43587     & HD 43587\\
                  &                       &              & corrected\\
\hline
Age [Gyrs]        & 2.17\p1.83            & 5.69\p0.57   & 4.97\p0.52\\
$M$ [M$_{\odot}$]  & 0.978\p0.026          & 1.037\p0.016 & 1.049\p0.016\\
$R$ [R$_{\odot}$]  & 1.00\p0.01            & 1.18\p0.03   & 1.15\p0.01\\
\logg \ [cgs]     & 4.46\p0.02            & 4.28\p0.01   & 4.30\p0.01\\
\hline
\end{tabular}
\tablefoot{The third column gives the PARAM results for HD 43587 Aa correcting for the presence of the companion.}
\end{table}

\begin{figure*}[t]
\centering
\includegraphics[width=15cm]{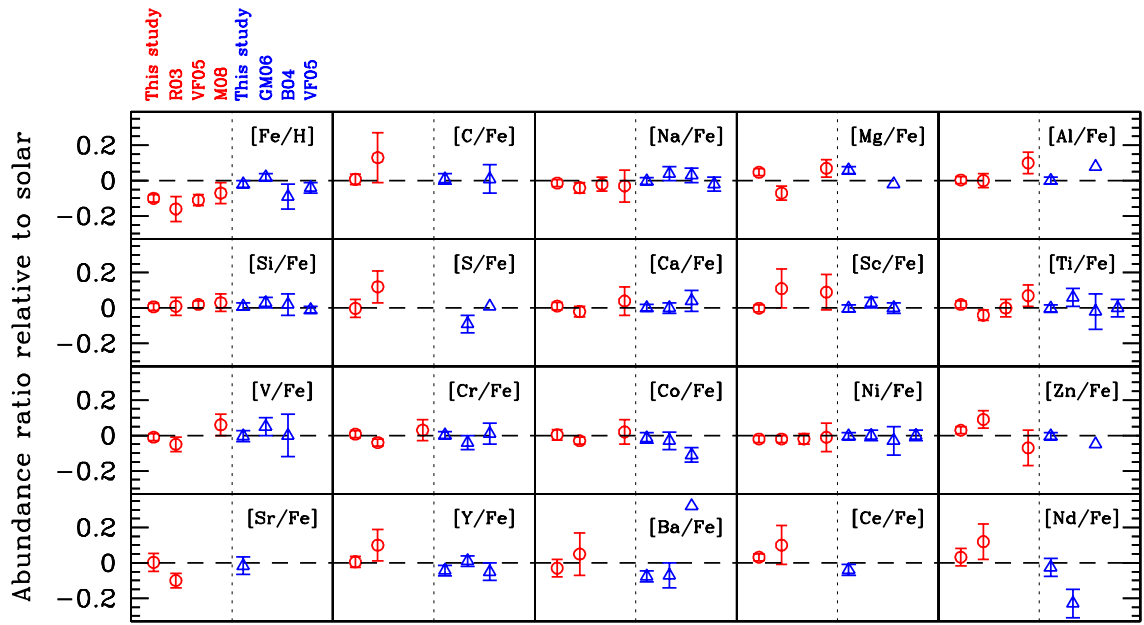}
\caption{Comparison between our abundances and previous results in the literature (R03: Reddy \etal \cite{reddy03}; VF05: Valenti \& Fischer \cite{valenti_fischer05}; M08: Mishenina \etal \cite{mishenina08}; GM06: Gillon \& Magain \cite{gillon_magain06}; B04: Bruntt \etal \cite{bruntt04}). The results for HD 42618 and HD 43587 are shown with circles and triangles, respectively. Our abundances based on a single spectral line have their uncertainty conservatively set to 0.05 dex.}
\label{fig_patterns_literature}
\end{figure*}

\subsection{Elemental abundances} \label{sect_reliability_abundances}
Our iron abundances are in accordance with those derived by previous studies, which range between $-$0.16 and $-$0.02 dex for HD 42618 and between $-$0.13 and +0.14 dex for HD 43587 (see references in caption of Fig.\ref{fig_teff_logg}). On the other hand, empirical calibrations between the strength of some near-infrared features and the metallicity yield for HD 43587B a \fe \ value fully consistent with that obtained here for the primary (e.g., Rojas-Ayala \etal \cite{rojas_ayala10}; Terrien \etal \cite{terrien12}). Our lithium abundance for HD 42618 agrees very well with the literature values (Mishenina \etal \cite{mishenina08}; Ram{\'{\i}}rez \etal \cite{ramirez12}; Takeda \etal \cite{takeda07}). Boesgaard \etal (\cite{boesgaard01}) obtained a lithium abundance lower by about 0.2 dex in HD 43587 ($\log \epsilon$[Li] = 1.84 dex) using parameters very similar to those determined here. A nearly identical value to that of Boesgaard \etal (\cite{boesgaard01}) was also obtained by Favata \etal (\cite{favata96}). However, their temperature based on photometric data (\teff \ = 5692 K) is significantly underestimated (Fig.\ref{fig_teff_logg}), and adopting a higher value would lead to a Li abundance closer to our estimate. Figure \ref{fig_patterns_literature} shows a comparison between our abundances and the results of the most comprehensive studies in the literature for these two objects (those of Takeda \etal \cite{takeda07} were ignored because no error bars are provided). Our values suggest an abundance pattern generally closer to that of the Sun in both cases. 

\section{Discussion} \label{sect_discussion}
As discussed previously, HD 42618 might be younger than the Sun (keeping in mind that ages are very uncertain for such unevolved field stars and that a very young age is not supported by the low activity level and lithium abundance), whereas HD 43587 is perhaps slightly older (Table \ref{tab_param}). The subsolar metallicity of HD 42618 considering its relative youth, if confirmed by asteroseismology in particular, might have developed because it was formed out of more metal-poor interstellar material in slightly outer parts of the Galaxy and migrated inwards afterwards to its current location in the solar vicinity (e.g., Haywood \cite{haywood08}). As a result of the possibly significantly different age of HD 42618 compared to the Sun, one might expect subtle differences in its abundance pattern because of the different relative proportion of Type Ia/II supernovae yields and the various amounts of material lost by stellar winds from asymptotic giant branch (AGB) and massive stars along the history of the Galaxy. Tentative (based on data for a dozen stars) relations between the stellar ages and the abundances relative to iron have recently been proposed for nearby, young dwarfs of the thin disk\footnote{As described in Morel \etal (\cite{morel03}), we used our own HARPS radial velocities and the new {\it Hipparcos} parallaxes and proper motions (van Leeuwen \cite{van_leeuwen07}) to compute the space components relative to the local standard of rest (LSR). We obtained ($U$, $V$, $W$)$_{\rm LSR}$ = (+52.3, --17.9, +3.8) and (--29.8, +9.6, --16.0) \kms \ for HD 42618 and HD 43587, respectively (a right-handed reference system, i.e., with $\vec{U}$ oriented towards the Galactic centre was adopted). We then followed Bensby \etal (\cite{bensby04}) to estimate the probability that the stars belong to a given kinematic component of the Galactic disk. We obtained that these two stars are 40 to 50 times more likely to belong to the thin than to the thick disk (similar figures have been obtained by Mishenina \etal \cite{mishenina04} and Ram{\'{\i}}rez \etal \cite{ramirez07}). Adopting another systemic velocity for HD 43587 Aa+b (the binary system has a semi-amplitude, $K$ = 4.3 \kms; Vogt \etal \cite{vogt02}) does not modify our conclusions.} by da Silva \etal (\cite{da_silva12}). However, the abundance pattern of HD 42618 appears to be very close to solar (among the elements with a determined uncertainty, magnesium is the most discrepant one but by only 2.5 $\sigma$; Fig.\ref{fig_patterns}). There is therefore no clear evidence that the abundances have been affected by the chemical evolution of the interstellar medium. Either the star is not much younger than the Sun and/or the differences are too small to be detected. 

\begin{figure}[h]
\centering
\includegraphics[width=9cm]{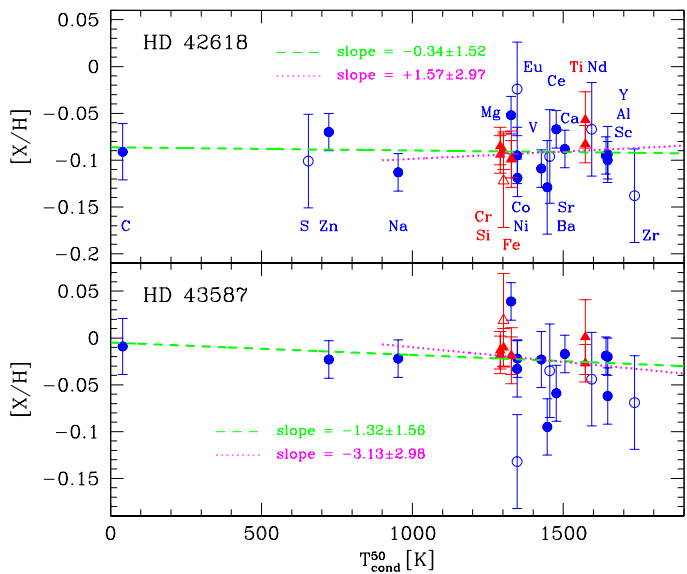}
\caption{Abundances as a function of the condensation temperature. Same symbols as in Fig.\ref{fig_patterns}. The fits weighted by the inverse variance, either taking into account all the elements or only those with $T^{50}_{\rm cond} >$ 900 K, are shown as dashed and dotted lines, respectively. The slopes (in units of 10$^{-5}$ dex K$^{-1}$) are indicated.}
\label{fig_condensation}
\end{figure}

Another phenomenon also potentially capable of producing slight departures from a scaled-solar chemical mixture is the formation of planetary companions. In particular, high-accuracy measurements have recently revealed small differences as a function of the condensation temperature between the solar abundances and those inferred in some solar twins/analogues. The different behaviour exhibited by volatile and refractory elements is believed to be related to the presence (or lack thereof) of terrestrial/gas giant planets (e.g., Mel\'endez \etal \cite{melendez09}). A search for planets orbiting our two targets using ultra-precise Doppler measurements gave negative results (Howard \etal \cite{howard10}; Vogt \etal \cite{vogt02}). As discussed above, the abundance pattern of HD 42618 is very close to the solar one and no trend between the abundance ratios [X/H] and the 50\% condensation temperature for a solar-photosphere composition gas $T^{50}_{\rm cond}$ (Lodders \cite{lodders03}) is discernible. A similar investigation for HD 43587 reveals that none of the slopes between [X/H] and $T^{50}_{\rm cond}$ (either over the whole $T^{50}_{\rm cond}$ range or over the domain sampled by the refractory elements: $T^{50}_{\rm cond}$ $\gtrsim$ 900 K) are statistically significant (see Fig.\ref{fig_condensation}). Flatter (but still compatible with zero) slopes are obtained when the heavy elements ($Z$ $>$ 30) that could bias the results because of chemical evolution effects are excluded (see Ram{\'{\i}}rez \etal \cite{ramirez11}). Flat slopes within the errors have also been reported for other solar analogues (e.g., Ram{\'{\i}}rez \etal \cite{ramirez09}; Gonz\'alez Hern\'andez \etal \cite{gonzalez_hernandez10}; \"Onehag \etal \cite{onehag11}). The abundances of several refractory elements are uncertain in HD 43587. If real, however, note that the global depletion relative to solar of these elements compared to the volatile ones hinted at by Fig.\ref{fig_condensation} is opposite to the behaviour generally observed in other solar analogues (e.g., Ram{\'{\i}}rez \etal \cite{ramirez09}).

\section{Summary and conclusion} \label{sect_conclusion}
Our study provides a number of important constraints, which will hopefully pave the way for a fruitful modelling of the \c \ data and robust inferences to be drawn about the internal structure of these two solar analogues. First, we provided an accurate determination of the atmospheric parameters and elemental abundances through a strictly differential analysis with respect to the Sun. While there is a remarkable agreement with the surface gravities derived from seismic data or isochrone fitting for HD 42618, this is less the case for HD 43587. The presence of a nearby, low-mass companion may contribute to these discrepancies, which remain small and comparable to the uncertainties, however. The abundance pattern of most metals relative to iron appears to be very close to solar for most chemical species (unlike what has been found for other solar-like \c \ targets; Bruntt \cite{bruntt09}), except for magnesium and a few trace elements in HD 43587, whose abundances are generally uncertain. On the other hand, solar abundances within the errors have been reported in these two targets for four elements not analysed here, namely O, K, Mn, and Cu (Bruntt \etal \cite{bruntt04}; Gillon \& Magain \cite{gillon_magain06};  Mishenina \etal \cite{mishenina08}; Ram{\'{\i}}rez \etal \cite{ramirez13}; Reddy \etal \cite{reddy03}). Assuming that the extent of microscopic diffusion is similar to that in the Sun, it is thus reasonable to assume a scaled-solar composition when modelling the \c \ data. Although we are unable to precisely determine their rotation rate, these stars are unlikely to be fast rotators unless they are seen from an unfavourable inclination angle (i.e., nearly pole on). Although not decisive in itself, their low activity levels seem to argue against this possibility. Generally speaking, the level of lithium depletion in the interior of main-sequence stars tends to increase with stellar metallicity and age, but to decrease at higher masses (e.g., Li \etal \cite{li12}). Detailed modelling is required to investigate whether the different Li surface abundances observed in our targets and in the Sun simply arise from their different fundamental parameters or instead reveal internal conditions diversely favourable to the destruction of this fragile element. A determination of the rotation properties and depth of the convective envelope from asteroseismology would be particularly valuable in this respect.

\begin{acknowledgements}
TM acknowledges financial support from Belspo for contract PRODEX GAIA-DPAC. MR and EP acknowledge financial support from the PRIN-INAF 2010 ({\it Asteroseismology: looking inside the stars with space- and ground-based observations}). We wish to thank F. Baudin and S. Mathur for providing preliminary results of the seismic analysis, and the referee, I. Ram{\'{\i}}rez, for his detailed report and useful suggestions that helped improve this paper. This research made use of NASA's Astrophysics Data System Bibliographic Services and the SIMBAD database operated at CDS, Strasbourg (France).
\end{acknowledgements}

\end{document}